\documentclass[12pt]{iopart}

\usepackage{graphicx,amssymb,bbm}

\def\tr{\,{\rm tr}\,}
\def\ket#1{|#1\rangle}
\def\bra#1{\langle#1|}
\def\braket#1#2{\langle #1 | #2 \rangle}
\def\bracket#1#2#3{\langle #1 | #2 | #3 \rangle}

\def\ii{\rm i}

\newcommand{\CC}{\mathbb{C}}
\def\dd{{\rm d}}
\def\PP{{\cal P}}

\begin{document}

\title{Detecting entanglement of random states with an entanglement witness}
\author{Marko \v Znidari\v c$^1$, Toma\v z Prosen$^1$, Giuliano Benenti$^{2,3}$ and Giulio Casati$^{2,3,4}$}
\address{${}^1$ Physics Department, Faculty of Mathematics and Physics,
University of Ljubljana, Jadranska 19, SI-1000 Ljubljana, Slovenia}
\address{${}^2$ CNISM, CNR-INFM \& Center for Nonlinear and Complex systems,
Universit\`{a} degli Studi dell'Insubria, via Valleggio 11,
I-22100 Como, Italy}
\address{${}^3$ Instituto Nazionale di Fisica Nucleare, Sezione di Milano,
via Celoria 16, I-20133 Milano, Italy}
\address{${}^4$ Department of Physics and Centre for Computational Science
and Engineering, National University of Singapore, Singapore 117542,
Republic of Singapore}

\begin{abstract}
The entanglement content of high-dimensional random pure states is almost maximal,
nevertheless, we show that, due to the complexity
of such states, the detection of their entanglement using witness operators is rather difficult.
We discuss the case of {\em unknown} random states, and the case of {\em known} random states  for which we can optimize the entanglement witness. Moreover, we show that coarse
graining, modeled by considering mixtures of $m$ random states instead of
pure ones, leads to a decay in the entanglement
detection probability exponential with $m$.
Our results also allow to explain the emergence of classicality in coarse grained quantum
chaotic dynamics.
\end{abstract}

\submitto{\JPA}
 
\section{Introduction}
\label{sec:intro}

Random pure quantum states, {\em i.~e.}, states whose distribution is invariant under an arbitrary unitary transformation,
are almost maximally entangled. Let us consider, for instance,
a random state $|\psi\rangle$ from $N \times N$ dimensional Hilbert space ${\cal H}=\CC^{N^2}$. We consider a bipartition to subsystems A and B, namely $\ket{\psi}$
as an element of
a tensor product Hilbert space ${\cal H}_{\rm A}\otimes {\cal H}_{\rm B}$ where
${\cal H}_{\rm A}=\CC^{N}$, ${\cal H}_{\rm B}=\CC^{N}$ are Hilbert spaces of subsystems
A and B, respectively.
The entanglement $E$ of the state $|\psi\rangle$ is then
measured by the von Neumann entropy
$S(\rho)=-\tr(\rho\log_2 \rho)$ of the reduced density matrices
$\rho_{\rm A}=\tr_{\rm B} |\psi\rangle\langle\psi|$ or
$\rho_{\rm B}=\tr_{\rm A} |\psi\rangle\langle\psi|$. It turns out
\cite{Lubkin:78,Lloyd:88,Page:93} that
\begin{equation}
E(|\psi\rangle)=S(\rho_{\rm A})
=S(\rho_{\rm B})\approx  \log_2 N -
\frac{1}{\log_e 4},
\end{equation}
which is close to the maximal value $\log_2 N$ attained for maximally
entangled state.
Since random states carry a lot of entanglement and
entanglement~\cite{plenio,entreview} has no analog in classical mechanics,
one can conclude that random states are highly non-classical.

On the other hand, pseudo-random states with properties close
to those of true random states can be efficiently generated by
dynamical systems (maps) in the regime of quantum chaos~\cite{stoeckmann,saraceno,schack,simone,caves,weinstein}.
In such chaotic maps the classical limit is recovered when
$N\to\infty$. Therefore, one can argue that for high-dimensional random states,
{\em i.e.}, in the limit $N\to \infty$, the quantum expectation value
of an operator with a well defined classical limit will be close to
its classical microcanonical average. According to this picture
random states in a way ``mimic'' classical microcanonical density.
Expectation values are therefore close to the classical ones.

At first sight this is in striking contrast with the almost maximal
entanglement of such states. How can we reconcile this apparent
contradiction? In the present paper we are going to tackle this question
by considering how can we detect entanglement of random states.
By studying the detection of entanglement with decomposable
entanglement witnesses we are going to argue that in the limit
of large systems the detection of entanglement in a random state
becomes increasingly difficult as it would demand the control of
very finely interwoven degrees of freedom and a measurement
resolution inversely proportional to the size of the Hilbert space,
which seems hardly feasible experimentally.
Therefore, as far as the detection of
entanglement is concerned, high dimensional random states are
effectively classical. 

Moreover, coarse graining naturally appears.
For instance, one could repeat several times the measurement of
the entanglement witness for a random state and the prepared
random state would be different from time to time due to
unavoidable experimental imperfections. We model this problem
by considering mixtures of $m$ pure random states, namely
\begin{equation}
\rho=\sum_{i=1}^m \frac{1}{m} \ket{\psi_i}\bra{\psi_i},
\label{eq:rho}
\end{equation}
where the $\ket{\psi_i}$ are mutually independent random pure states, but in general
they are not orthogonal. We are going to show that
the detection of entanglement is even more difficult for these
mixed states, as it requires a number of measurements growing
exponentially with $m$.

There are other physical contexts in which formally the same kind of coarse graining (mixing of
the state) naturally appears:  (i) {\em Time averaging.} For example, if a state $\ket{\psi}$ undergoes
a time evolution $\ket{\psi(t)} = U(t)\ket{\psi}$ given in terms of some unitary dynamics $U(t)$,
then the time average of a physical observable $A$ over an interval $T$ is essentially determined
by expectation value $\tr(A \rho)$ in the mixed state 
\begin{equation}
\rho=\frac{1}{T}\int_0^T {\rm d} t \ket{\psi(t)}\bra{\psi(t)}
\label{eq:rhot}
\end{equation} 
which has an effective rank $m \approx T/t_{\rm corr}$ where $t_{\rm corr}$ is a dynamical correlation
time of the dynamics $U(t)$. For a quantum chaotic evolution $U(t)$, the state $\ket{\psi(t)}$ can be,
after some time, arguably well described by a random state and the correlation time $t_{\rm corr}$ is expected to be short,  so $\rho$ \,(\ref{eq:rhot}) may be considered as a  mixture (\ref{eq:rho}) 
of $m$ uncorrelated random states.
(ii) {\em Phase space averaging.} Sometimes  it is useful to represent quantum states
 in terms of distribution functions in the classical phase space, like the Husimi function 
 (see e.g. \cite{saraceno}), which can
be understood as a convolution of the Wigner function or its coarse graining over a 
 phase space volume $(2\pi\hbar)^d$ in $d$ degreees of freedom.
 In fact, the Husimi function of a pure state can be understood as a Wigner function of
 the following mixed state 
 \begin{equation}
 \rho = (2\pi\hbar)^{-d} \int {\rm d}\vec{q}{\rm d}\vec{p} 
 \exp\left(-\frac{1}{2\hbar}(\alpha q^2+\alpha^{-1} p^2)\right) T(\vec{q},\vec{p})\ket{\psi}\bra{\psi}T^\dagger(\vec{q},\vec{p})
 \label{eq:rhop}
 \end{equation} 
 where $T(\vec{q},\vec{p})$ are unitary phase space translation operators, and $\alpha$ is an arbitrary squeezing parameter. A random pure state $\ket{\psi}$ has a Wigner function with random sub-Planck structures with phase space correlaton length $l_{\rm corr} \sim \hbar$ which is semi-classically smaller than the
 coarse-graining width $\sim \hbar^{1/2}$, so $\rho$ (\ref{eq:rhop}) can be again considered as a mixture  (\ref{eq:rho}) of $m$ random pure states with $m \sim \hbar^{-1/2}$.
 
The paper is organized as follows. In Sec.~\ref{sec:ew}, we
review known results about entanglement witnesses and
pure states entanglement. Depending on our a priori knowledge
of the random state, two relevant cases can be distinguished:
(i) We do not know in advance of which random state
we are going to detect the entanglement, {\em i.e.}, the
random state is unknown.
In such a case the best one can do is to choose some generic
entanglement witness in advance, independently of the state
we measure. Such situation can also be thought to arise in
the case when we are not able to prepare an arbitrary entanglement
witness but just some subset of witnesses.
(ii) We know the random state in advance and we are able
to prepare an arbitrary entanglement witness. In such case we can
use the optimal entanglement witness for each random state separately.
These two cases are discusses in Secs.~\ref{sec:urs} and
~\ref{sec:krs}, respectively. Finally, in Sec.~\ref{sec:conc}, we
provide a brief discussion of our results.

\section{Entanglement witnesses}
\label{sec:ew}

First of all, let us introduce some known facts about entanglement
witnesses and entanglement of pure states that we will need for
the derivation of our results in subsequent sections.
Given a pure state $\ket{\psi}$, its bipartite entanglement
content is completely specified by the Schmidt decomposition:
\begin{equation}
\ket{\psi}=\sum_i \mu_i \ket{a_i} \otimes \ket{b_i},
\label{eq:Sch}
\end{equation}
where the Schmidt coefficients $\mu_i$ are positive real numbers
satisfying $\sum_i \mu^2_i=1$ and $\ket{a_i}$ and $\ket{b_i}$ are
orthonormal states on subspaces ${\cal H}_{\rm A}$ and
${\cal H}_{\rm B}$.  The squares of the Schmidt coefficients are equal
to the eigenvalues of the reduced density matrices $\rho_{\rm A}$
and $\rho_{\rm B}$. For such bipartition there are at most $N$ nonzero
Schmidt coefficients $\mu_i$. For a state having $r$ roughly equal nonzero
Schmidt coefficients ($\mu_i\sim 1/\sqrt{r}$),
the reduced von Neumann entropy is $S(\rho_A)$
$\sim \log_2{r}$. For instance, the simple (GHZ) state
$\ket{\psi}=\frac{1}{\sqrt{2}}\ket{0\ldots0}+\frac{1}{\sqrt{2}}
\ket{1\ldots 1}$ leads to two nonzero eigenvalues of $\rho_{\rm A}$
($\mu_1=\mu_2=1/\sqrt{2}$), and simple Schmidt vectors
$|a_1\rangle\otimes|b_1\rangle = \ket{0\ldots0}$ and
$|a_2\rangle\otimes|b_2\rangle = \ket{1\ldots 1}$, resulting in $S=1$.
On the other hand, a random pure state has all $N$ eigenvalues
nonzero, very complicated (random) Schmidt vectors $\ket{a_i}$ and
$\ket{b_i}$, and almost maximal entropy
$S\approx \log_2 N-1/\log_e{4}$~\cite{Lubkin:78,Lloyd:88,Page:93}.
Note, however, that all the eigenvalues of the reduced density
matrix for a bipartite random state are very small.
In fact, they decrease with $N$, the largest being, on average, $4/N$ while the
smallest is $1/N^3$~\cite{JPA07}. The guiding idea of this paper is that
this smallness of eigenvalues and the complexity of eigenvectors should
be somehow reflected in the difficulty of detecting entanglement
in such a state, despite the fact that the entanglement content of a
random state is large.

Besides pure states, we will also be interested in the entanglement of a mixture of pure states (\ref{eq:rho}).
A quantum state $\rho$ of a bipartite system is called separable if it can be written as
\begin{equation}
\rho=\sum_k p_k\,  \rho_{{\rm A}k} \otimes
\rho_{{\rm B}k},\quad
p_k\ge 0,\quad
\sum_k p_k=1,
\end{equation}
where $\rho_{{\rm A}k}$ and $\rho_{{\rm B}k}$ are density
matrices for the two subsystems. A state is entangled if it is not separable. 
To decide whether a given mixed state is entangled
or not is a difficult problem \cite{plenio,entreview}.
Fortunately, there is an operational criterion which is able to
detect the most useful entangled states. This is the famous positive
partial transposition (PPT) criterion~\cite{PPT}: since separable
states have positive semidefinite partial transpose\footnote{Introducing
and orthonormal basis $\{\ket{i\alpha}=|i\rangle_{\rm A}\otimes|\alpha\rangle_{\rm B}\}$
in the Hilbert space associated with the bipartite system, the
density matrix $\rho$ has matrix elements
$\rho_{i\alpha,j\beta}=\bra{i\alpha}\rho\ket{j\beta}$.
The partial transpose
is constructed by taking the transpose of only Latin or only Greek
indexes (here Latin indexes refer to subsystem A and
Greek indexes to subsystem B). For instance, the partial transpose
with respect to subsystem B is given by
\begin{equation}
\rho^{\rm T_B}_{i\alpha,j\beta}=
\rho_{i\beta,j\alpha}.
\end{equation}}, all non-PPT
states are entangled. Note, however, that for dimensions higher
than $2\times 2$ and $2\times 3$ there exist PPT-entangled states,
known as bound entangled states \cite{W}.

A convenient way to detect entanglement is to use
the so-called entanglement witnesses~\cite{W,Terhal:00}.
By definition, an entanglement witness is a Hermitian operator $W$
such that $\tr(W\rho_{\rm sep})\ge 0$ for all separable states $\rho_{\rm sep}$
while there exists at least one state $\rho_{\rm ent}$ such that
$\tr(W\rho_{\rm ent})<0$. Therefore, the negative expectation value
of $W$ is a signature of entanglement and the state $\rho_{\rm ent}$
is said to be detected as entangled by the witness $W$.
The concept of entanglement witness is close to experimental
implementations and detection of entanglement by means of
entanglement witnesses has been realized in several
experiments~\cite{Bour:04,wineland,Haffner:05}.
Because it is easier to measure a larger negative expectation value of an entanglement witness, one can argue that the detection of entanglement is easier the larger this negative value is. The expectation value of $W$ also provides lower bounds to various entanglement
measures~\cite{eisert:07,Guhne:07}. Estimation of entanglement entropy by measurement of observables has been considered in~\cite{Klich:06}.
Finally, it is interesting to note that violation of Bell inequalities
can be rewritten in terms of nonoptimal entanglement
witnesses~\cite{Terhal:00,Hyllus:05}.

In general, classification of entanglement witnesses is a hard problem. However,
much simpler is the issue with the so-called decomposable entanglement
witnesses (D-EW)~\cite{Lewenstein:00}. D-EW is a witness which
can be decomposed as
\begin{equation}
W=P+Q^{\rm T_B}, \qquad P,Q\ge 0,
\label{eq:DEW}
\end{equation}
that is, with positive semidefinite operators $P,Q$. D-EW can detect only
non-PPT entangled states, {\em i.e.}, those with negative eigenvalues
of $\rho^{\rm T_B}$. They are therefore equivalent to the PPT criterion
but closer to experimental implementations, as full tomographic
knowledge about the state is not needed. In the present paper
we are going to limit ourselves only to D-EW. General non-decomposable
entanglement witness (ND-EW) can be written in a canonical form
as $W=P+Q^{\rm T_B}-\epsilon
\mathbbm{1}$~\cite{Lewenstein:00,Lewenstein:01} and can also detect
entangled states with PPT. Finding an optimal ND-EW, for which violation
of positivity is maximal, is in general hard~\cite{Lewenstein:00}.

\section{Unknown random state}
\label{sec:urs}

In this section we assume that the random state $\ket{\psi}$ whose
entanglement we would like to detect is unknown so that we are not
able to use an optimal $W$ for a particular $\ket{\psi}$. The best
one can do is to choose some fixed entanglement witness $W$ in advance,
independently of the state. Since we will be interested in the average
behaviour over unitary invariant ensemble of pure random states, $W$
can be chosen to be random as well. That is, in the present section
we are going to study detection of entanglement with a random
entanglement witness, whose precise definition will be given later.
What we want to calculate is the distribution of the expectation
values $\bracket{\psi}{W}{\psi}$ for a fixed $W$ and an ensemble of
random pure states $\ket{\psi}$. Averaging over random states
$\ket{\psi}$ we see that the average expectation value
$\overline{\bracket{\psi}{W}{\psi}}$ is
\begin{equation}
\overline{\bracket{\psi}{W}{\psi}}=\int{\dd\PP \bracket{\psi}{W}{\psi}}=\tr{W}/N^2,
\end{equation}
where $\overline{\bullet} = \int{\dd\PP}\bullet$ denotes an integration over a unique $U(N^2)$-invariant distribution
of pure states  $\ket{\psi}$, and we used the fact that for
a random state $\ket{\psi}=\sum_i c_i \ket{i}$
we have $\overline{c_i c_j^*}=\delta_{ij}/N^2$.
We fix normalization of the entanglement witness $W$ such that
$\tr W=1$. Therefore, the average expectation value
$\overline{\bracket{\psi}{W}{\psi}}$ scales $\propto 1/N^2$.
We therefore define the rescaled quantity
$w=N^2 \bracket{\psi}{W}{\psi}$, such that $\overline{w}=1$,
independently of the dimension $N$. From now on we will focus on the random variable $w$,
and its distribution with density $p(w) = \dd\PP/\dd w$.

Because operator $P$ in D-EW (\ref{eq:DEW}) just shifts the
expectation value towards positive values, we limit
ourselves to D-EW of the form $W=Q^{\rm T_B}$.
Any positive semidefinite operator $Q$ can be written in its eigenbasis in terms of
positive eigenvalues $d_i$, satisfying $\sum_i d_i = 1$, and eigenvectors $\ket{\phi_i}$, hence
$W=\sum_i d_i (\ket{\phi_i}\bra{\phi_i})^{\rm T_B}$. We will first
study the case when $Q$ is a simple rank one projector (subsection \ref{subsect:Qproj}), that is $W$
is given by $W=(\ket{\phi}\bra{\phi})^{\rm T_B}$. There are two reasons
why this is the most important case. First, as we will see below,
optimal D-EW is always of such a ``projector'' form. Second, the expectation
value for a general $Q$ can be written (subsection \ref{subsect:Qany}) as a sum of expectation values
for individual eigenvectors of $Q$ and therefore the probability distribution
of $w$ will be a simple convolution of distributions for the case of
$Q$ being rank one.

Most of our theoretical results are derived for one mixing component only, $m=1$ (i.e.
for a pure random state),  since the general result for arbitrary number $m$ of independent 
mixing components is obtained by simple convolutions as discussed in subsection 
\ref{subsect:anym}.
  
\subsection{Q is a projector}
\label{subsect:Qproj}

The entanglement witness is of the form
$W=(\ket{\phi}\bra{\phi})^{\rm T_B}$, with a fixed $\ket{\phi}$, and we
would like to calculate the density of probability distribution $p(w)$ of its normalized expectation
values $w$ for random pure states. This distribution can depend
on the chosen $\ket{\phi}$. First few moments of the density
$p(w)$ can be calculated explicitly. We have already seen
that $\overline{w}=1$. For higher moments we can in the
leading order (in Hilbert space dimension $1/N^2$) use Gaussian
averages and Wick contractions in order to approximate integrals over the
unitary group \cite{weiden}.
Using
\begin{equation}
\overline{c_i^* c_j c_k^* c_l}=
(\delta_{ij}\delta_{kl}+\delta_{il}\delta_{jk})/N^4 + {\cal O}(N^{-6}),
\end{equation}
and similarly for higher order products, we arrive, up to corrections ${\cal O}(N^{-2})$, at
\begin{eqnarray}
\kappa_2 &=& \overline{(w-\overline{w})^2}=\tr{W^2}=1, \nonumber\\
\kappa_3 &=& \overline{(w-\overline{w})^3}=2\tr{W^3}, \label{eq:trW}\\
\kappa_4 &=& \overline{(w-\overline{w})^4}-3\kappa_2^2=6 \tr{W^4}. \nonumber
\end{eqnarray}
Here we denote by $\kappa_n$ the $n$-th cumulant.
We see that the average value of $w$ as well as the width of the distribution
is $1$, independently of the state $\ket{\phi}$ we use for $W$.
While $\tr{W}=1$ is a simple normalization, second moment
is $1$ due to $Q$ being of rank $1$. If only the first two cumulants were
nonzero our probability density $p(w)$ would be a simple Gaussian.

\begin{figure}[!h]
\centerline{\includegraphics[width=80mm,angle=-90]{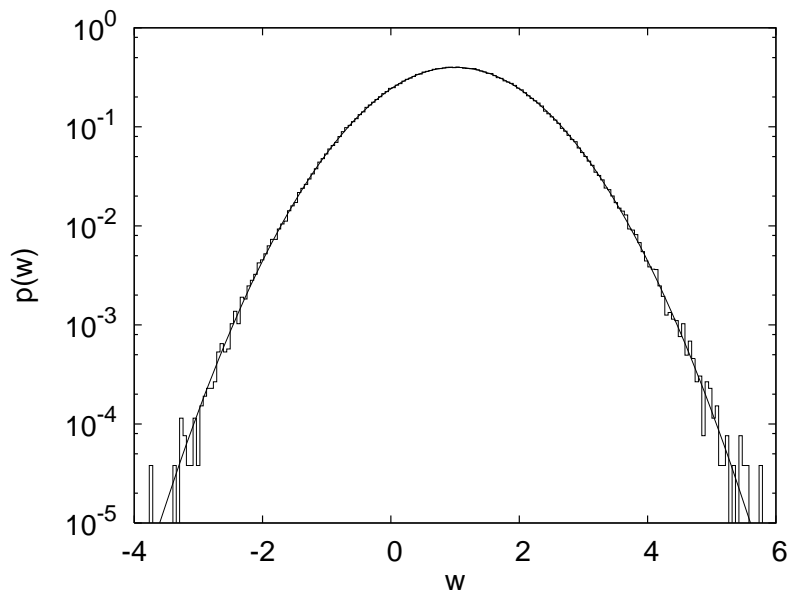}}
\caption{Density of probability distribution of $w=N^2\bracket{\psi}{W}{\psi}$ for random
vectors $\ket{\psi}$ and a single $W=(\ket{\phi}\bra{\phi})^{\rm T_B}$
with a random $\ket{\phi}$. Total Hilbert space dimension $N^2=2^{10}$.
Full curve is the theoretical Gaussian prediction (\ref{eq:gauss}). We use an ensemble of $5\times 10^5$ random states.}
\label{fig:gauss}
\end{figure}

Let us first see what happens if we choose for $\ket{\phi}$ a state
with large Schmidt rank $r$ (number of nonzero eigenvalues of the
reduced density matrix $\sigma_{\rm A}=\tr_{\rm B}\ket{\phi}\bra{\phi} $) of order
$r \sim N$. Using Schmidt decomposition (\ref{eq:Sch})
for the state $\ket{\phi}=\sum_i \mu_i \ket{a_i}\otimes \ket{b_i}$ we can immediately write eigenvalues and eigenvectors
of the Hermitian operator $W=(\ket{\phi}\bra{\phi})^{\rm T_B}$. There are
$N(N-1)$ eigenvalues $\pm \mu_i \mu_j$, $i < j$, with the corresponding
eigenvectors $(\ket{a_i b_j^*}\pm \ket{a_j b_i^*})/\sqrt{2}$ and
$N$ eigenvalues $\mu_i^2$ with the corresponding eigenvectors
$\ket{a_i b_i^*}$. In our notation, $\ket{a_i b_j^*}=\ket{a_i}\otimes\ket{b_j^*}$, where 
$\ket{b_i^*}=\sum_\alpha b_{i\alpha}^* \ket{\alpha}$ if $\ket{b_i} = \sum_\alpha b_{i\alpha}\ket{\alpha}$.
Using these eigenvalues one can see that
the traces of powers of $W$ are 
\begin{equation}
\tr{W^{2k}}=(\tr{\sigma_{\rm A}^k})^2,\qquad
\tr{W^{2k+1}}=\tr{\sigma_{\rm A}^{2k+1}}.
\end{equation}
Because eigenvalues of
$\sigma_{\rm A}$ are of order $\sim 1/r$, we have $\tr{W^3} \sim 1/r^2
\sim {\cal O}(1/N^2)$, $\tr{W^4} \sim 1/r^2 \sim {\cal O}(1/N^2)$ and so
on. Note that formulas (\ref{eq:trW}) are exact up to order $1/N^2$
and therefore can not be used for $\kappa_{n\ge 3}$ and states
$\ket{\phi}$ with the full Schmidt rank. Nevertheless, we can conclude that
the higher cumulants are at most $\sim 1/N^2$ and therefore vanish in the limit $N \to \infty$.
Therefore, for $W$ and $\ket{\phi}$ with an increasing rank,
the probability density $p(w)$ converges to a Gaussian in the limit $N\to\infty$,
\begin{equation}
p(w)=\frac{1}{\sqrt{2\pi}}\exp{(-(w-1)^2/2)}.
\label{eq:gauss}
\end{equation}
Theoretical prediction (\ref{eq:gauss}) is compared with the
results of numerical simulation in figure~\ref{fig:gauss}.
Probability to measure negative $w$, {\em i.e.}, of detecting
entanglement, is $\int_{-\infty}^0{p(w)dw}$ and therefore
\begin{equation}
\PP(w<0)=(1-{\rm erf}(1/\sqrt{2}))/2 \approx 0.159.
\end{equation}
Note that this entanglement detection probability is independent
of the details of $\ket{\phi}$, provided that its Schmidt
rank $r$ is large, more precisely $r\propto N$.

The above Gaussian form of $p(w)$ can be understood also as a
consequence of the central limit theorem. Indeed, let us write
$W$ in its eigenbasis, then we have, writing $\lambda_i=\mu_i^2$,
\begin{equation}
w=\sum_{i=1}^{r} \lambda_i
|\braket{\psi}{a_i b_i^*}|^2+\sum_{i<j} \sqrt{\lambda_i \lambda_j}
(\braket{\psi}{a_i b_j^*}\braket{a_j b_i^*}{\psi}+
\braket{\psi}{a_j b_i^*}\braket{a_i b_j^*}{\psi}).
\end{equation}
Denoting overlaps by $\braket{\psi}{a_i b_j^*}=\sqrt{y_{ij}}
{\rm e}^{\ii \varphi_{ij}}$, where $y_{ij}$ and $\varphi_{ij}$
are two real numbers, we have
\begin{equation}
w=\sum_{i,j=1}^{r}\sqrt{\lambda_i \lambda_j}\sqrt{y_{ij} y_{ji}}
\cos{(\varphi_{ij}-\varphi_{ji})}.
\label{eq:w}
\end{equation}
For the overlap of two random states we know that the angle $\varphi$
is distributed uniformly, while the amplitude (scaled by $N^2$) has
an exponential distribution, $p(y_{ij})=\exp{(-y_{ij})}$. Therefore,
for a given set of eigenvalues $\lambda_i$ of $\sigma_{\rm A}$, and assuming
$y_{ij}$ to be independent, we can calculate the distribution of
$w$ using the above formula. It is a convolution of distributions of
individual terms. In the limit $r \to \infty$ the central limit theorem
can be used, resulting in a Gaussian distribution (\ref{eq:gauss}).

\begin{figure}[!h]
\centerline{\includegraphics[width=80mm,angle=-90]{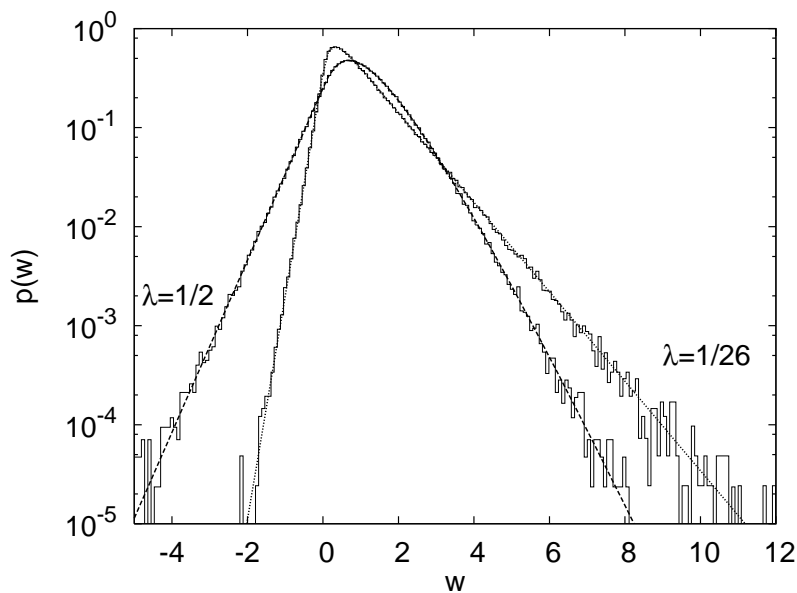}}
\caption{Density of probability distribution of $w=N^2\bracket{\psi}{W}{\psi}$ for random
vectors $\ket{\psi}$ and a single $W=\ket{\phi}\bra{\phi}^{\rm T_B}$
with $\ket{\phi}$ having two nonzero Schmidt coefficients
$\sqrt{\lambda}$ and $\sqrt{1-\lambda}$. We show two cases,
$\lambda=1/2$ and $\lambda=1/26$, both for $N^2=2^{10}$. Dotted curve
is theoretical prediction (\ref{eq:pw_sch}) for $\lambda=1/26$ while
the dashed one is for $\lambda=1/2$, see eq.~(\ref{eq:pw_sch2}).
We use an ensemble of $5\times 10^5$ random states.}
\label{fig:sch}
\end{figure}

For $\ket{\phi}$ with a small finite Schmidt rank 
there will be deviations
from Gaussian because higher cumulants of $p(w)$ (\ref{eq:trW}) will
be in general nonzero also in the limit $N\to\infty$. The above form
for $w$ (\ref{eq:w}) is actually very handy for an explicit calculation
of $p(w)$ for $\ket{\phi}$ with a small rank $r$. Let us take the extreme case of
rank $r=2$, where we expect strongest deviations from a Gaussian.
We therefore have only two nonzero eigenvalues of $\sigma_{\rm A}$,
$\lambda_1=\lambda$ and $\lambda_2=1-\lambda$. Assuming the overlaps
$y_{ij}$ to be independent and exponentially distributed and the angles
$\varphi_{ij}$ to be uniform, we arrive after evaluating few
convolutions at
\begin{equation}
\fl
p(w)=
\cases{
  \frac{1}{(1-2\lambda)^2} \left\{\lambda {\rm e}^{-\frac{w}{\lambda}} + (1-\lambda){\rm e}^{-\frac{w}{1-\lambda}} \right\}+ \frac{1}{4\sqrt{\lambda(1-\lambda)}-2}{\rm e}^{-\frac{w}{\sqrt{\lambda(1-\lambda)}}}\!\!\!\!\!\!\!\!\! & : $w>0$, \cr
  \frac{1}{4\sqrt{\lambda(1-\lambda)}+2}{\rm e}^{\frac{w}{\sqrt{\lambda(1-\lambda)}}}\!\!\!\!\!\!\!\!\!  & : $w<0$.
}
\label{eq:pw_sch}
\end{equation}
The distribution is a sum of exponentials. Probability to detect entanglement,
{\em i.e.}, to measure negative value of $w$ is
$\PP(w<0)=1/(4+2/\sqrt{\lambda(1-\lambda)})$.
As a function of $\lambda$ it reaches a maximum for $\lambda=1/2$,
{\em i.e.}, both eigenvalues of $\sigma_{\rm A}$ are equal, where it is
equal to $1/8$ (which is less than for Gaussian
distribution (\ref{eq:gauss})). In figure~\ref{fig:sch} we compare results
of numerical simulation for $p(w)$ for two cases: one with $\lambda=1/2$,
for which the appropriate limit of eq.~(\ref{eq:pw_sch}) gives
\begin{equation}
p(w)_{\lambda=1/2}=
\cases{
  (1+4w+8w^2)\frac{{\rm e}^{-2w}}{4} & : $w>0$, \cr
  \frac{1}{4}{\rm e}^{2w}  & : $w<0$.
}
\label{eq:pw_sch2}
\end{equation}
and the case with an almost pure $\sigma_{\rm A}$, $\lambda=1/26$, for which the
probability of detecting entanglement is $5/72\approx 0.07$. Note that in
the limit $\lambda \to 0$, {\em i.e.}, of a pure product state for
$\ket{\phi}$, $w$ will always be positive with an exponential distribution.

\subsection{Q of higher rank}
\label{subsect:Qany}

So far we have discussed only the case when $Q$ is a one-dimensional projector,
$Q=\ket{\phi}\bra{\phi}$. What happens if the rank of $Q$ is larger?
If $Q$ is of rank $2$, $W$ can be written as $W=d_1 W_1+d_2 W_2$ with
positive $d_{1,2}$ and $W_{1,2}=(\ket{\phi_{1,2}}\bra{\phi_{1,2}})^{\rm T_B}$.
If we assume $W_1$ and $W_2$ are statistically independent,
so that $w_1=N^2 \bracket{\psi}{W_1}{\psi}$ and $w_2=N^2
\bracket{\psi}{W_2}{\psi}$ are uncorrelated, then the distribution of
$w=d_1 w_1+d_2 w_2$ is given by a convolution of distributions for
$w_{1,2}$. Let us calculate the second moment of $w$ given by the above sum.
Using 
\begin{equation}
N^2 \int \dd\PP \bracket{\psi}{W_1}{\psi}\bracket{\psi}{W_2}{\psi}=
\tr{(W_1 W_2)}+\tr{W_1}\tr{W_2}+{\cal O}(1/N),
\end{equation}
and
\begin{equation}
\tr{(W_1 W_2)}=|\braket{\phi_1}{\phi_2}|^2,
\end{equation} 
we get
\begin{equation}
\overline{(w-\overline{w})^2}=d_1^2+d_2^2+2d_1 d_2 |\braket{\phi_1}{\phi_2}|^2.
\end{equation}
We see that the width of the distribution of $w$ is the same as in
the case of convolution of two independent distributions,
leading to a width $d_1^2+d_2^2$, provided the two vectors
$\ket{\phi_1}$ and $\ket{\phi_2}$ are orthogonal.
Using similar considerations
for higher moments we can conclude that in the case of
$Q=\sum_i^k d_i \ket{\phi_i}\bra{\phi_i}$ having rank $k$, and 
because eigenvectors of $Q$ are orthogonal, the distribution of $w$
is Gaussian of width $\sigma^2=\sum_i^k d_i^2$, at least for
sufficiently large $k$. Assuming for simplicity that all
$d_i$ are the same, $d_i=1/k$, we obtain
\begin{equation}
p(w)=\sqrt{\frac{k}{2\pi}} {\rm e}^{-k(w-1)^2/2}.
\label{eq:pwr}
\end{equation}

\subsection{Mixtures of random states}
\label{subsect:anym}

It is also interesting to consider the case in which $Q$ is of rank $1$
but we wish to detect the entanglement of mixed states, for instance
of a mixture of $m$ pure random states, as given in eq.(\ref{eq:rho}).
In such case the expectation value is
$w=\frac{1}{m}\sum_i\tr({W \ket{\psi_i}\bra{\psi_i}})$, with
$W=(\ket{\phi}\bra{\phi})^{\rm T_B}$. Repeating the same derivation
as for $Q$ of higher rank and pure states, just replacing $k$ by $m$, 
we see that the distribution of $w$ will be a convolution of distributions for individual
$\ket{\psi_i}$ and, provided these are statistically independent, the width of the
resulting Gaussian (\ref{eq:pwr}) will be $1/m$.
Because the probability density $p(w)$ becomes narrowly peaked about its
mean $\overline{w}=1$ with increasing $m$, the probability to measure
negative values decreases with $m$, namely for the Gaussian form of
eq.~(\ref{eq:pwr}) we have
\begin{equation}
\PP(w<0)=\frac{1-{\rm erf}(\sqrt{m/2})}{2}\asymp
\frac{1}{\sqrt{2\pi m}} {\rm e}^{-m/2}.
\label{eq:pm}
\end{equation}
This probability decays to zero exponentially with $m$.
Therefore, the detection of entanglement for a mixture of random states
is very hard.

\section{Known random state}
\label{sec:krs}

In this section we assume that the random state $\ket{\psi}$ whose
entanglement we want to measure is known in advance and furthermore,
that we are able to prepare an arbitrary D-EW. 
In addition, we have to assume that our state $\ket{\psi}$ is neither separable, nor
bound entangled, which is true with probability which converges to one exponentially
in $N$. Therefore, for
each $\ket{\psi}$ we can prepare an optimal entanglement witness,
such that its expectation value will be minimal. As far as D-EW
are concerned, the optimal choice of $W=W_{\rm opt}$ is to take for $Q$
a projector to the eigenspace corresponding to the minimal
(negative) eigenvalue $\lambda_{\rm min}$ of $\rho^{\rm T_B}$,
$W_{\rm opt}=(\ket{\phi_{\rm min}}\bra{\phi_{\rm min}})^{\rm T_B}$.
The maximal violation of positivity is therefore
\begin{equation}
\tr{(W_{\rm opt}\rho)}=-|\lambda_{\rm min}(\rho^{\rm T_B})|.
\label{eq:Wopt}
\end{equation}
If we are able to measure entanglement witness with a given precision\footnote{Different
normalization of $W$, {\em e.g.}, $\tr{W}=f(N)$, would result in the
maximal violation $-f(N)|\lambda_{\rm min}|$.}
it is the size of $\lambda_{\rm min}$ which determines the difficulty
of detecting entanglement in $\ket{\psi}$. Note that the optimal
entanglement witness $W_{\rm opt}$ depends on the state $\ket{\psi}$.
For each state $\ket{\psi}$ we have to pick a different $W_{\rm opt}$.

\subsection{Distribution of eigenvalues of $\rho^{\rm T_B}$}

First, let us look at the distribution of eigenvalues after PT operation
for a single random pure state. The eigenvalues can be written in terms of Schmidt coefficients $\mu_i$ as
$\lambda=\pm \mu_i \mu_j$. Formally, we can write the distribution of $\lambda$, for $\lambda > 0$, as
\begin{equation}
\frac{\dd\PP}{\dd\lambda} = \int\!\dd\mu_i \int\!\dd\mu_j \,\delta(\lambda - \mu_i\mu_j) 
\frac{\dd\PP}{\dd\mu_i}  \frac{\dd\PP}{\dd\mu_j} 
\end{equation}
where distribution of $\mu_i$ can be for large $N$ obtained from
Mar\v cenko-Pastur law~\cite{Marcenko:67} for
the distribution of $\tau = N \mu_i^2$, namely 
\begin{equation}
\dd\PP/\dd\tau  = \sqrt{\tau(4-\tau)}/(2\pi\tau),
\end{equation}
by a simple change of variables $\tau \to \mu_i$.
The result for the distribution of scaled eigenvalues
$y=N\lambda$ reads
\begin{equation}
\frac{\dd\PP}{\dd y} =\frac{1}{8 \pi^2}\left[ (16+y^2)K(1\!-\!y^2/16)-32
E(1\!-\!y^2/16) \right],\quad y \in [-4,4],
\label{eq:py}
\end{equation}
where $E(x)$ and $K(x)$ are elliptic integrals. Note that this distribution
takes into account only the $N(N-1)$ ``off-diagonal'' eigenvalues
$\pm \mu_i \mu_j$ with $i < j$. ``Diagonal'' eigenvalues $\mu_i^2$ have
the same distribution as the eigenvalues of $\rho_{\rm A}$ and are only
$N$ in number. The expectation value of the minimal eigenvalue equals 
$\overline{\lambda}_{\rm min}=-4/N$. In fact the distribution of $\lambda_{\rm min}$ becomes
strongly peaked around $-4/N$ with diminishing relative fluctuations as $N\to\infty$. 

\begin{figure}[!h]
\centerline{\includegraphics[width=55mm,angle=-90]{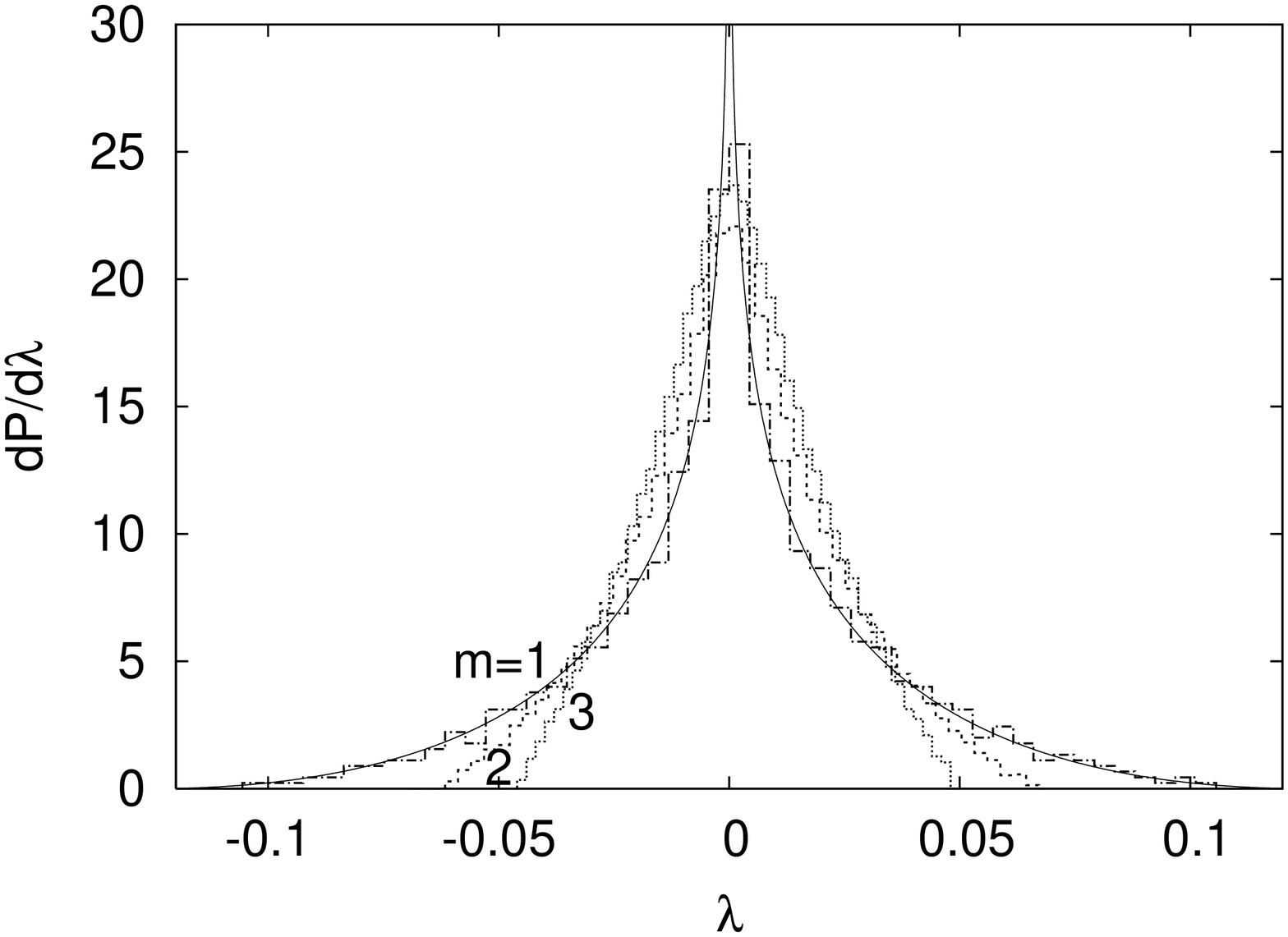}\includegraphics[width=55mm,angle=-90]{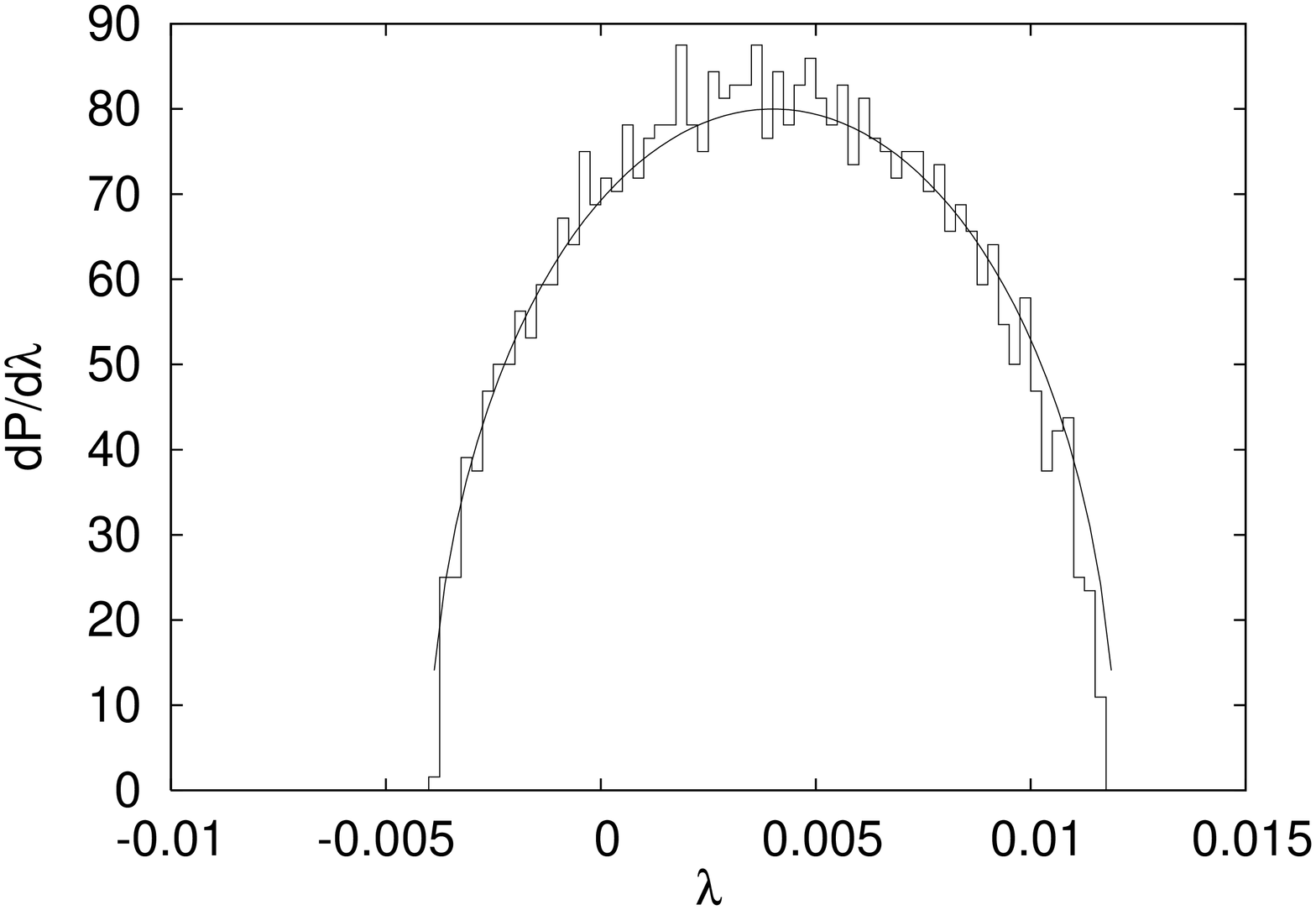}}
\caption{Distribution of eigenvalues of $\rho^{\rm T_B}$, eq.~(\ref{eq:rho}),
for mixing few states, $m=1,2,3$ (left plot), all for $N^2=2^{10}$.
Theoretical formula for $m=1$ (\ref{eq:py}) is shown with the full curve. The right frame
shows the distribution after mixing many states, $m=256$, $N^2=2^8$. Full curve
is a semicircle with the center at $1/N^2$. All histograms are averages over an ensemble of $10$ 
mixtures $\rho$ (\ref{eq:rho}).
}
\label{fig:hist10_U0}
\end{figure}
When we mix several independent (in general non-orthogonal) random vectors,
 $\rho=\sum_i^m \ket{\psi_i}\bra{\psi_i}/m$, the minimal eigenvalue
$\lambda_{\rm min}$ increases and the distribution becomes increasingly
sharply peaked (for $m \to \infty$ we get $\rho \to \mathbbm{1}/N^2$
with all eigenvalues being equal to $1/N^2$). We numerically verified
the above theoretical prediction for $\dd\PP/\dd y$ (\ref{eq:py})
in figure~\ref{fig:hist10_U0}.
We can also see that after mixing many random states ($m \sim N$) the distribution
becomes a semicircle, which is a numerical result for which we have yet no analytical explanation.

\subsection{Minimal eigenvalue of $\rho^{\rm T_B}$}

\begin{figure}[!h]
\centerline{\includegraphics[width=55mm,angle=-90]{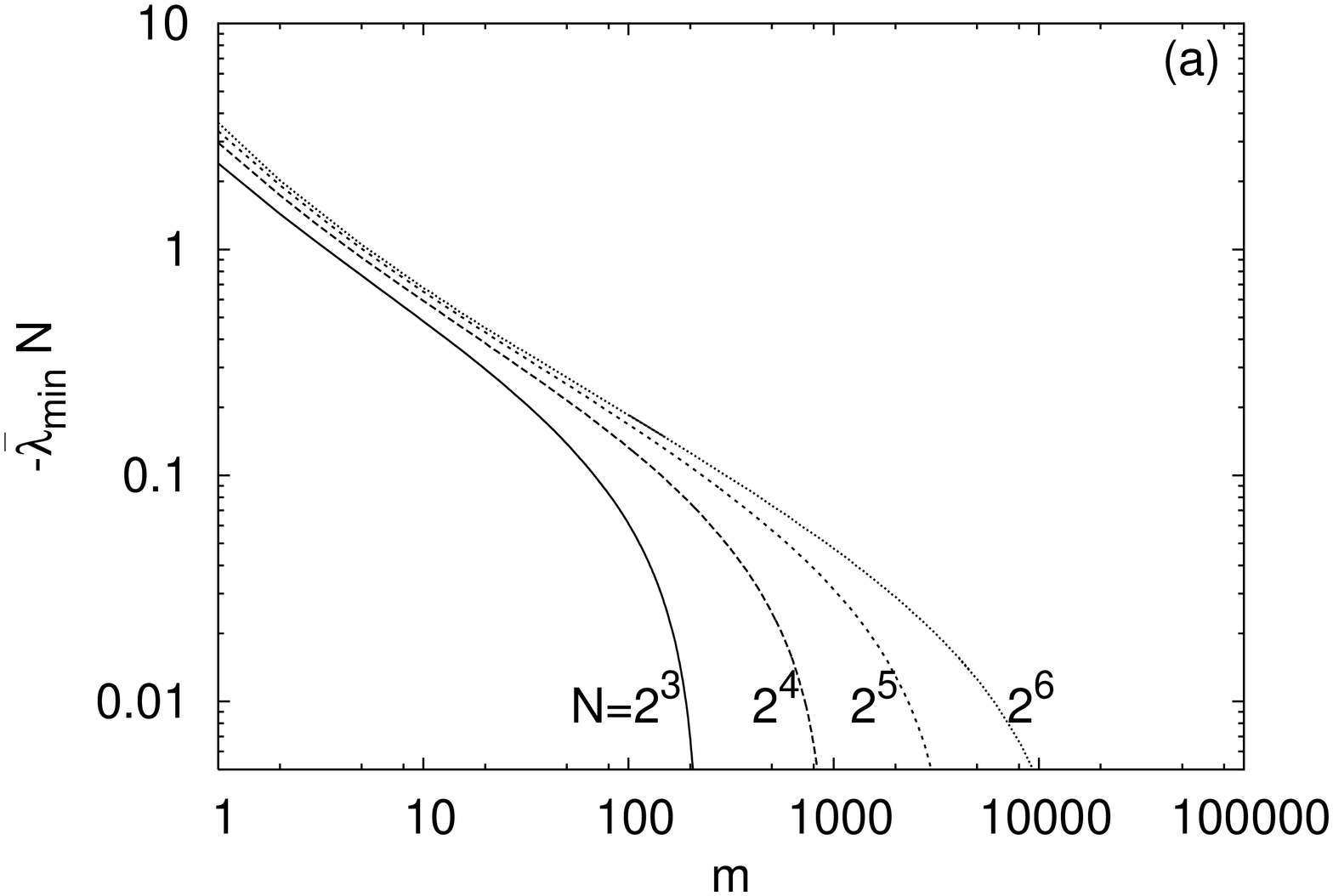}\includegraphics[width=55mm,angle=-90]{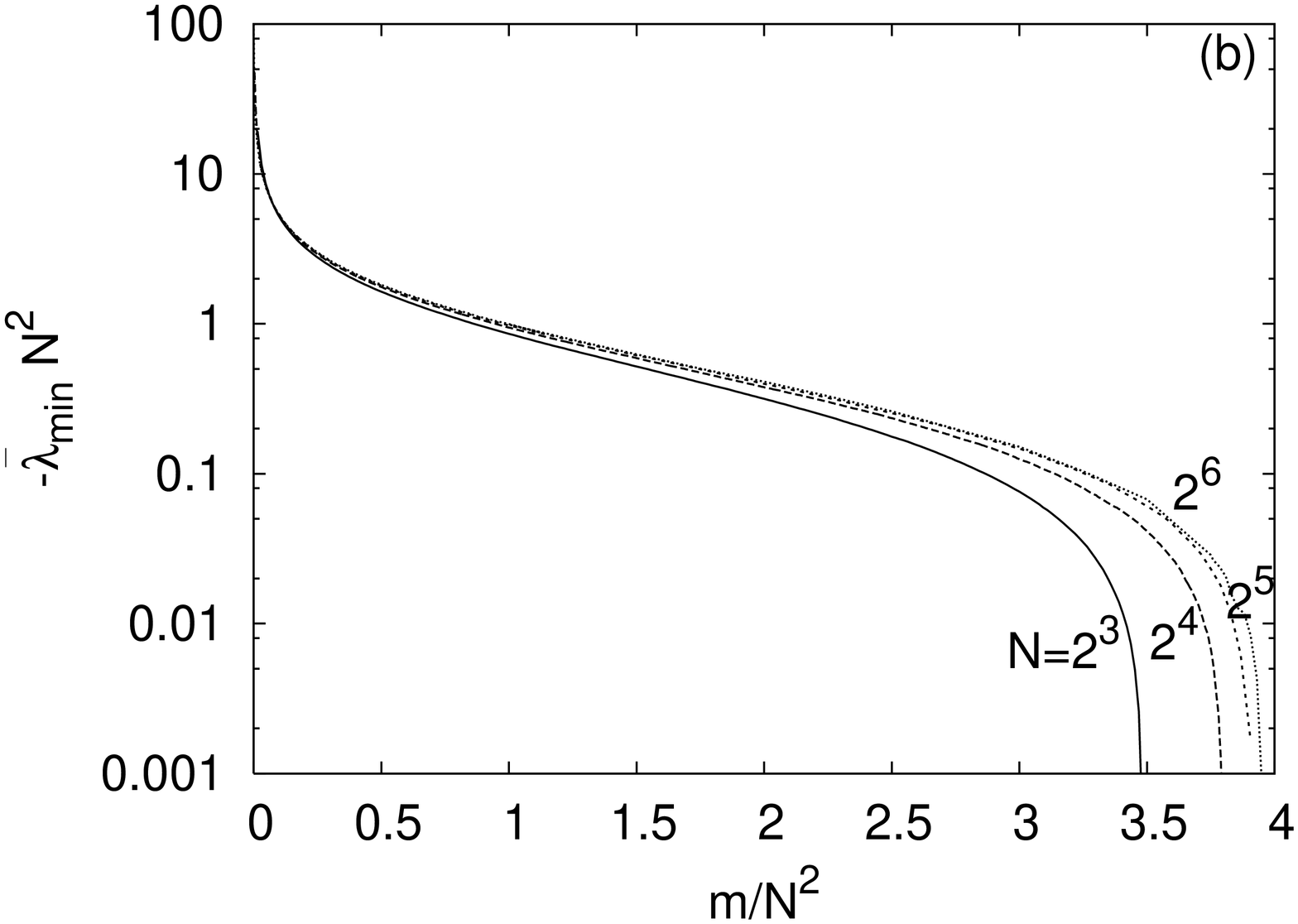}}
\caption{Dependence of $\overline{\lambda}_{\rm min}$ on $m$. Overlapping of curves
for different dimension $N$ signals the scaling $\lambda_{\rm min}\sim -1/N$ for small
$m$ in the left frame (a) and $\lambda_{\rm min}\sim -1/N^2$ for large $m$ (but smaller than $m^*$ where $\bar{\lambda}_{\rm min}$ changes sign) in the
right frame (b). We average over $1000,100,10$, and $6$ mixed random states (\ref{eq:rho}),
for $N^2=2^6,2^8,2^{10}$ and $2^{12}$, respectively.
}
\label{fig:lmin}
\end{figure}

Because $\lambda_{\rm min}$ determines the maximal violation for an optimal
witness $W$ (\ref{eq:Wopt}) we are going to look in more detail at the
dependence of $\lambda_{\rm min}$ on $m$ and $N$. For a single pure state
($m=1$) we know that the average minimal eigenvalue is $\overline{\lambda}_{\rm min}=-4/N$.
On the other hand, we also know that for large $m$, as we approach completely mixed state the minimal eigenvalue must scale as $\overline{\lambda}_{\rm min} \sim 1/N^2$. Therefore, the scaling of
$\overline{\lambda}_{\rm min}$ has to change as we increase the number of mixed
states $m$. To confirm this expectation, we performed numerical simulation,
calculating the average $\overline{\lambda}_{\rm min}$ for different $m$. The results
are shown in figure~\ref{fig:lmin}. Note that the average minimal eigenvalue $\bar{\lambda}_{\rm min}$ is positive for $m > m^*$, with $m^* \approx 4 N^2$.

Although von Neumann entropy of a random state is large all eigenvalues
of $\rho^{\rm T_B}$ are very small and will therefore be hard to detect.
If we assume that we are able to measure values of $\tr{(W\rho)}$ with 
accuracy $\epsilon$ then we can, 
depending on the scaling of $\epsilon$ with $N$, tell for which
values of $m$ the detection of entanglement is possible. If $\epsilon$ does not
depend on $N$, {\em i.e.}, precision does not increase with $N$, then for
sufficiently large $N$, such that $4/N <\epsilon$, detection of entanglement
will be impossible. Already a single random state becomes from the viewpoint
of entanglement detection ``classical'', since measuring a negative expectation value
of its optimal entanglement witness is below the detection limit. If on
the other hand we are able to measure $\epsilon$ which decreases as
$1/N$, the critical $m_{\rm crit}$, beyond which the entanglement
detection is impossible, will be independent of $N$, {\em i.e.}, in the
limit $N\to\infty$ the ratio $m_{\rm crit}/N \to 0$ (see fig.~\ref{fig:lmin}a). If however we are able
to detect very small expectation values of order $1/N^2$, then
$m_{\rm crit}$ will be proportional to $N^2$ (see fig.~\ref{fig:lmin}b).
Furthermore, even with arbitrary accuracy, detection of entanglement with D-EW is impossible
beyond $m=m^* \propto N^2$.

Worth mentioning is that for $m \ge N$ the mixed state
$\rho$ of eq.(\ref{eq:rho})
can not be used for dense coding. Indeed a quantum state is useful
for dense coding~\cite{dense} only if $S(\rho_{\rm A})-S(\rho)>0$
(see {\em e.g.}~\cite{DC}). For mixtures of $m$ random states
$S(\rho)$ is roughly equal to $\sim \log m$. On the other hand
$S(\rho_{\rm A})$ is at most $\log{N}$. Therefore, $S(\rho_{\rm A})$
will be smaller than $S(\rho)$ for $m \sim N$.

\section{Conclusions}
\label{sec:conc}

In this paper we have considered random states and have shown that,
while their entanglement content is almost maximal, the detection
of such entanglement is very difficult. This is a consequence of
the complexity of a random state, which leads to a large number
of small coefficients in the Schmidt decomposition of the state.
Nevertheless, for random pure states, a finite success probability
in the detection of entanglement exists also in the limit in which
the Hilbert space dimension $N^2\to\infty$. This implies that quantum
chaos alone is not sufficient to erase any trace of entanglement
in the classical limit, provided that ideal measurements are possible.
On the other hand such erasure becomes
very efficient when coarse graining is taken into account, for
instance when mixtures instead of pure states are considered.

We note that all our results can be straightforwardly generalized to the
case of unbalanced bipartition $N={\rm dim}{\cal H}_{\rm A}\neq N'={\rm dim}{\cal H}_{\rm B}$,
for instance, distribution of $w = N N' \bra{\psi}W\ket{\psi}$ is Gaussian
(\ref{eq:gauss}) provided both dimensions $N,N'$ are large.

We would like to stress once more than the detection
difficulties are a consequence of the complexity of random states.
If instead one considers ``regular states'' such as the GHZ state
$|{\rm GHZ}\rangle=\frac{1}{\sqrt{2}}
(|0...0\rangle+|1...1\rangle)$, then the optimal
witness is $W_{\rm opt}=(|\phi_{\rm min}\rangle\langle\phi_{\rm min}|)^{\rm T_B}$,
with $|\phi_{\rm min}\rangle=
\frac{1}{\sqrt{2}}(|0...0\rangle|1...1\rangle
-|1...1\rangle|0...0\rangle)$ which corresponds to the minimal eigenvalue
$\lambda_{\rm min}=-1/2$ of $(|{\rm GHZ}\rangle
\langle{\rm GHZ}|)^{\rm T_B}$.
Since the value of $\lambda_{\rm min}$ is $-1/2$ instead of $-4/N$
as for a random state, it turns out that it will be much easier
to detect entanglement in a ``regular'' rather than in a random
state. This happens in spite of the fact that the entanglement 
content is larger in a random than in such a regular state.

\ack
The authors would like to thank Slovenian Research Agency, programme P1-0044,
grant J1-7437, and the MIUR-PRIN 2005 (2005025204) for support.


\section*{References}

\begin{thebibliography}{1}

\bibitem{Lubkin:78} E.~Lubkin, {\em Entropy of an n-system from its correlation with a k-reservoir}, J.~Math.~Phys. {\bf 19}, 1028-1031 (1978).

\bibitem{Lloyd:88} S.~Lloyd and H.~Pagels, {\em Complexity as thermodynamic depth}, Ann.~Phys.~(NY) {\bf 188}, 186-213 (1988).

\bibitem{Page:93} D.~Page, {\em Average entropy of a subsystem}, Phys.~Rev.~Lett. {\bf 71}, 1291 (1993).

\bibitem{plenio} M.~B.~Plenio and S.~Virmani, {\em An introduction to
entanglement measures}, Quant.~Inf.~Comp. {\bf 7}, 1 (2007).

\bibitem{entreview} R.~Horodecki, P.~Horodecki, M.~Horodecki and K.~Horodecki, {\em Quantum entanglement}, quant-ph/0702225.

\bibitem{stoeckmann} H.-J.~ St\" ockmann, {\em Quantum Chaos: An introduction},
(Cambridge University Press, Cambridge 1999).

\bibitem{saraceno}
M.~Saraceno, {\em Classical structures in the quantized baker transformation},
Ann.~Phys. (NY) {\bf 199}, 37 (1990).

\bibitem{schack} 
R.~Schack, {\em Using a quantum computer to investigate quantum chaos}, Phys.~Rev. A {\bf 57}, 1634 (1998).

\bibitem{simone} G.~Benenti, G.~Casati, S.~Montangero and D.~L.~Shepelyansky, {\em Eigenstates of an operating quantum computer: hypersensitivity
to static imperfections}, Eur.~Phys.~J.~D {\bf 20}, 293 (2002).

\bibitem{caves} A.~J.~Scott and C.~M.~Caves, {\em Entangling power of
the quantum baker's map}, J.~Phys.~A {\bf 36}, 9553 (2003).

\bibitem{weinstein}
Y.~S.~Weinstein and C.~S.~Hellberg, {\em Entanglement generation of nearly random operators}, Phys.~Rev.~Lett. {\bf 95}, 030501 (2005).

\bibitem{JPA07} M.~\v Znidari\v c, {\em Entanglement of random vectors}, J.~Phys.~A {\bf 40}, F105 (2007).

\bibitem{PPT} A.~Peres, {\em Separability criterion for density matrices}, Phys.~Rev.~Lett. {\bf 77}, 1413 (1996).

\bibitem{W} M.~Horodecki, P.~Horodecki and R.~Horodecki, {\em Separability of mixed states: necessary and sufficient conditions}, Phys.~Lett.~A {\bf 223}, 1 (1996).

\bibitem{Terhal:00} B.~Terhal, {\em Bell inequalities and the separability criterion}, Phys.~Lett.~A {\bf 271}, 319 (2000).

\bibitem{Bour:04} M.~Bourennane, M.~Eibl, C.~Kurtsiefer, S.~Gaertner, H.~Weinfurter, O.~G\" uhne, P.~Hyllus, D.~Bru\ss, M.~Lewenstein and Anna Sanpera, {\em Experimental detection of multipartite entanglement using witness operators}, Phys.~Rev.~Lett. {\bf 92}, 087902 (2004).

\bibitem{wineland} D.~Leibfried, E.~Knill, S.~Seidelin, J.~Britton,
R.~B.~Blakestad, J.~Chiaverini, D.~B.~Hume, W.~M.~Itano, J.~D.~Jost,
C.~Langer, R.~Ozeri, R.~Reichle, and D.~J.~Wineland,
{\em Creation of a six-atom 'Schr\"odinger cat' state},
Nature {\bf 438}, 639 (2005).

\bibitem{Haffner:05} H.~H\" affner, W.~H\" ansel, C.~F.~Roos, J.~Benhelm, D.~Chek-al-kar, M.~Chwalla, T.~K\" orber, U.~D.~Rapol, M.~Riebe, P.~O.~Schmidt, C.~Becher, O.~G\" uhne, W.~D\" ur and R.~Blatt, {\em Scalable multi-particle entanglement of trapped ions}, Nature {\bf 438}, 643 (2005).

\bibitem{eisert:07} J.~Eisert, F.~G.~S.~L. Brandao and K.~M.~R. Audenaert, {\em Quantitative entanglement witnesses}, New J.~Phys. {\bf 9}, 46 (2007).

\bibitem{Guhne:07}  O.~G\" uhne, M.~Reimpell and R.~F.~Werner, {\em Estimating entanglement measures in experiments}, Phys.~Rev.~Lett. {\bf 98}, 110502 (2007).

\bibitem{Klich:06} I.~Klich, G.~Refael and A.~Silva, {\em Measuring entanglement entropes in many-body systems}, Phys.~Rev.~A {\bf 74}, 032306 (2006).

\bibitem{Hyllus:05} P.~Hyllus, O.~G\" uhne, D.~Bru\ss and M.~Lewenstein, {\em Relations between entanglement witnesses and Bell inequalities}, Phys.~Rev.~A {\bf 72}, 012321 (2005).

\bibitem{Lewenstein:00} M.~Lewenstein, B.~Kraus, J.~I.~Cirac and P.~Horodecki, {\em Optimization of entanglement witnesses}, Phys.~Rev.~A {\bf 62}, 052310 (2000).

\bibitem{Lewenstein:01} M.~Lewenstein, B.~Kraus, P.~Horodecki and J.~I.~Cirac, {\em Characterization of separable states and entanglement witnesses}, Phys.~Rev.~A {\bf 63}, 044304 (2001).

\bibitem{weiden} T.~Prosen, T.~H.~Seligman and H.~A.~Weidenm\" uller, {\em Integration over matrix spaces with unique invariant measures},  J. Math. Phys. {\bf 43},  5135 (2002).

\bibitem{Marcenko:67} {V.~A.}~{Mar\v cenko} and {L.~A.}~Pastur, {\em Distribution of eigenvalues of some sets of random matrices}, Math.~USSR-Sb {\bf 1}, 457-483 (1967).

\bibitem{dense} C.~H.~Bennett and S.~J.~Wiesner, {\em Communication via one- and two-particle operators on Einstein-Podolsky-Rosen states}, Phys.~Rev.~Lett. {\bf 69}, 2881 (1992).

\bibitem{DC} A.~Sen, U.~Sen, M.~Lewstein and A.~Sanpera, {\em The separability versus entanglement problem}, quant-ph/0508032.


\end{thebibliography}
\end{document}